\documentclass[a4paper,twocolumn, 9pt]{article}
\usepackage[paperwidth=210mm,paperheight=297mm,centering,hmargin=2.3cm,vmargin=2.7cm]{geometry}
\usepackage{marvosym}
\usepackage{amsfonts}
\usepackage{amsthm}
\usepackage{dcolumn}
\usepackage{graphicx}
\usepackage{bm,bbm}
\usepackage{kpfonts}
\usepackage{braket}
\usepackage{color}

\newcommand{\cH}{{\mathcal H}}

\newcommand{\cK}{{\mathcal K}}
\newcommand{\cL}{{\mathcal L}}

\newcommand{\cN}{{\mathcal N}}

\newcommand{\cU}{{\mathcal U}}
\newcommand{\cV}{{\mathcal V}}

\newcommand{\cX}{{\mathcal X}}

\newcommand{\fT}{{\mathfrak T}}

\newcommand{\bbmC}{{\mathbbm C}}

\newcommand{\bbmN}{{\mathbbm N}}
\newcommand{\bbmR}{{\mathbbm R}}

\newcommand{\id}{\mathrm{id}}

\newcommand{\conv}{\mathrm{conv}}

\newtheorem{theorem}{Theorem}

\newtheorem{definition}[theorem]{Definition}

\newtheorem{lemma}[theorem]{Lemma}

\newtheorem{proposition}[theorem]{Proposition}
\newtheorem{remark}[theorem]{Remark}

\newcommand{\U}{\mathcal U}

\begin{document}
\title{Randomness cost of symmetric twirling} 
\author{Holger Boche, Gisbert Jan\ss en, Sajad Saeedinaeeni\\
\scriptsize{Electronic addresses: \{boche, gisbert.janssen, sajad.saeedinaeeni\}@tum.de}
\vspace{0.2cm}\\
{\footnotesize Lehrstuhl f\"ur Theoretische Informationstechnik, Technische Universit\"at M\"unchen,}\\
{\footnotesize 80290 M\"unchen, Germany }}
\maketitle

\begin{abstract}
We study random unitary channels which reproduce the action of the twirling channel corresponding to the representation of the symmetric group on an $n$-fold tensor product. We derive upper
and lower bounds on the randomness cost of implementing such a map which depend exponentially on the number of systems. Consequently, symmetric twirling can be regarded as a reasonable Shannon-theoretic protocol. On the other hand, such protocols are disqualified by their resource-inefficiency in situations where randomness is a costly resource.
\end{abstract}
\section{Introduction} \label{sect:introduction}
When designing communication protocols, the quantum information theorist has a vast and steadily growing toolbox of approved protocol parts at hand. Especially useful are universal protocols, which perform a certain task regardless of the preparation of the system. \newline
As a prominent example of this class we mention the quantum teleportation protocol which allows noiseless transmission of an unknown qubit state by using a pure maximally entangled qubit pair and two bits of noiseless forward communication. The fact that the teleportation protocol perfectly accomplishes this goal is completely independent of the state to be transmitted, motivates modular use in larger protocols without further adjustment of the protocol. In this paper, we address symmetric twirling, which perfectly and universally transforms each state on a given $n$-party system to a permutation invariant one. This is accomplished by applying a unitary $U^{\pi}$ which exchanges the subsystems according to a permutation $\pi$ which is chosen randomly according to the equidistribution on the group $S_n$ of permutations on $n$ elements, i.e. the quantum channel
\begin{align*}
\U(\cdot) := \frac{1}{|S_n|}\sum_{\pi \in S_n} U^{\pi}(\cdot){U^{\pi}}^{\dagger}.
\end{align*}
is applied. \newline
This protocol is very useful in in situations where the system is demanded to be permutationally invariant for further processing. An example of such a situation is, where $\cU$ is performed to make a system ready for applying an instance of the quantum de Finetti Theorem (see for example \cite{als}.) 
While application of $\cU$ makes all states on the underlying systems perfectly permutation-invariant, 
the protocol is highly demanding regarding its randomness cost. Since $n!$ grows superexponentially with the number $n$ of systems, the randomness cost of the protocol is not bounded by any rate. This fact prevents $\cU$ from being a reasonable protocol in situations where randomness is at all counted as a resource. However, the equidistributed choice out of all permutations obviously bares some redundancies, such that a less randomness consuming way of choosing permutations to emulate $\cU$ seems possible.\newline In Section \ref{sect:symmetric_designs} of this paper, we derive upper and lower bounds on the randomness needed to perform the symmetric twirling channel $\cU$, both of which lie on the exponential scale. We also show, that the lower bound essentially remains valid under the weakened condition, that the action of the twirling is simulated only approximately well. 
\newline
In Section \ref{sect:applications}, we discuss the consequences of our findings for communication theory. The upper bounds derived show that the action of symmetric twirling indeed can be accomplished universally by a protocol with rate-bounded randomness demands. This fact is important in situations where randomness is not a free resource (e.g. when the random permutations have to be applied by two or more parties in a coordinated way.) On the other hand, the lower bounds derived show, that the randomness needed is close to the maximum randomness which can be generated from that system. 
Therefore, symmetric twirling is too expensive in some situations. Such situations arise especially, when the random choice of permutations has to be kept private from additional adversarial communication parties. \newline
\emph{Related work:}
The twirling and determination of randomness needed to perform such was extensively studied in the work \cite{evenly} in case of the group of unitary transformations on a given Hilbert space. Therein, the notion of a \emph{unitary design} was introduced, a terminology which we extend to the symmetric group in this work. Recently, Wakakuwa \cite{wakakuwa16} determined the asymptotic randomness cost of symmetrizing a given quantum state in case of tensor product representations of an arbitrary given group. We point out, that the focus set in \cite{wakakuwa16} is different from this work. Namely, the twirling we consider does not arise from a tensor product representation and is therefore out of the scope of \cite{wakakuwa16}. Moreover, we are focused on protocols which emulate the twirling operation \emph{universally}, while the mentioned work rather asks for the randomness cost of simulating the action of a twirling for a fixed state.
\section{Bounds for symmetric designs} \label{sect:symmetric_designs}
Considerations briefly explained in the introduction, have brought up the question of whether it is possible to render the average over a group using only a subset of its elements. In particular, $S_{n}$, the group of all permutations on $n$ elements (or equivalently all bijections over $\{1,...,n\}$), will be of interest in the remainder of our work. We consider the unitary representation $\{U^{\pi}\}_{\pi\in S_{n}}$ of this group acting on $(\mathbb{C}^{d})^{\otimes n}$, defined by the action:
\begin{equation}
	U^{\pi} x_{1}\otimes...\otimes x_{n}=x_{\pi(1)}\otimes...\otimes x_{\pi(n)}
\end{equation}
for each $\pi\in S_{n}$ and $x_{1}, \dots, x_{n}\in\mathbb{C}^{d}$. We prove bounds for all weighted subsets of $S_{n}$, the unitary representations of which produce the group's average. We refer to such subsets as \textit{symmetric weighted designs}, as they are the analogous objects to spherical or unitary designs (\cite{konig},\cite{evenly}).
\begin{definition} \label{def:weighted_symmetric_design}
	Let $X\subset S_{n}$ and $\omega:X\to\mathbb{R}^{+}$ be a weight function (i.e. $\omega>0$ and $\sum_{\pi\in X}\omega(\pi)=1$). The pair $(X,\omega)$ is a symmetric weighted design (or a weighted design for $S_{n}$), if:
	\begin{equation}\label{definition}
		\frac{1}{n!}\sum_{\pi\in S_{n}}U^{\pi}\eta {U^{\pi}}^{\dagger}=\sum_{\pi\in X}\omega(\pi)U^{\pi}\eta {U^{\pi}}^{\dagger}
	\end{equation}
	for all $\eta\in\mathcal{L}(\mathcal{H}^{\otimes n})$, where $\mathcal{H}:=\mathbb{C}^{d}$. 
\end{definition}
To prove the upper bound on the cardinality of the designs we use the following theorem from convex analysis. A proof can be found in e.g. \cite{barvinok03}, Theorem 2.3.
 \begin{theorem}[Carath\'{e}odory's Theorem]
 Let $S \subset \bbmR^d$ be a set. Then every point $x \in \conv(S)$ can be represented as a convex combination of $d+1$ points from $S$, i.e. there exist $\alpha_1,\dots,\alpha_{d+1} \geq 0$, $\sum_{i=1}^{d+1} \alpha_i = 1$, and $y_1,\dots,y_n \in S$ such that
 \begin{align}
  x = \alpha_1 y_1 + \dots + \alpha_{d+1} y_{d+1}
 \end{align}
 holds.
 \end{theorem}
 
\begin{theorem} \label{thm:symmetric_design_upperbound}
	There exists a symmetric weighted design $(X,\omega)$ with cardinality of $X$ upper-bounded as:
		\begin{equation}\label{upperbound}
		|X|\leq d^{4n}+1
	\end{equation}
\end{theorem}
\begin{proof}
Let $B:=\{\ket{e_{x}}:x\in[d]\}$ be the standard basis for $\mathbb{C}^{d}$. We will use the notation 
\begin{align}
 \ket{e_{\mathbf{x}}} := \ket{e_{x_1}} \otimes \cdots \otimes\ket{e_{x_n}}  
\end{align}
for each $\mathbf{x} := (x_1,\dots,x_n) \in \cX^{n}$.
Writing the left and right hand sides of (\ref{definition}) in terms of matrix entries of $U^{\pi}$ and ${U^{\pi}}^{\dagger}$ (${u^{\pi}}^{\dagger}_{ij}:\bra{e_{i}}{U^{\pi}}^{\dagger}\ket{e_{j}},u^{\pi}_{ij}:\bra{e_{i}}U^{\pi}\ket{e_{j}}$) we obtain:
\begin{align*}
	L:=\frac{1}{n!}\sum_{\pi\in S_{n}}U^{\pi}\eta {U^{\pi}}^{\dagger}=
\end{align*}

\begin{align}
	\frac{1}{n!}\sum_{\pi\in S_{n}}\sum_{\mathbf{w},\mathbf{x},\mathbf{y},\mathbf{z}\in[d^{n}]} a_{\mathbf{x}\mathbf{y}}u^{\pi}_{\mathbf{w}\mathbf{x}}{u^{\pi}}^{\dagger}_{\mathbf{y}\mathbf{z}}\ket{e_{\mathbf{w}}}\bra{e_{\mathbf{z}}}
\end{align}
and
\begin{equation}
	R:=\sum_{\pi\in X}\omega(\pi)\sum_{\mathbf{w},\mathbf{x},\mathbf{y},\mathbf{z}\in[d^{n}]}a_{\mathbf{x}\mathbf{y}}u^{\pi}_{\mathbf{w}\mathbf{x}}{u^{\pi}}^{\dagger}_{\mathbf{y}\mathbf{z}}\ket{e_{\mathbf{w}}}\bra{e_{\mathbf{z}}}
\end{equation}
where $a_{\mathbf{x}\mathbf{y}}:=\bra{e_{\mathbf{x}}}\eta\ket{e_{\mathbf{y}}}$. Since $a_{\mathbf{x}\mathbf{y}}$ only depends on $\eta$, it can be observed that $R=L$ (and hence $(X,\omega)$ is a symmetric weighted design) if we have:
\begin{equation}\label{conditiondesign}
	\frac{1}{n!}\sum_{\pi\in S_{n}}u^{\pi}_{\mathbf{w}\mathbf{x}}{u^{\pi}}^{\dagger}_{\mathbf{y}\mathbf{z}}=\sum_{\pi\in X}\omega(\pi)u^{\pi}_{\mathbf{w}\mathbf{x}}{u^{\pi}}^{\dagger}_{\mathbf{y}\mathbf{z}}
\end{equation}
for all $\mathbf{w},\mathbf{x},\mathbf{y},\mathbf{z}\in[d^{n}]:=\{1,...,d\}^{n}$. Define the vector $\nu_{\pi}:=(u^{\pi}_{\mathbf{w}\mathbf{x}}{u^{\pi}}^{\dagger}_{\mathbf{y}\mathbf{z}}:\mathbf{w},\mathbf{x},\mathbf{y},\mathbf{z}\in[d^{n}])$. We observe that $\nu_{\pi}\in{\mathbb{R}^{d}}^{4n}$, as the entries of $U^{\pi}$ are either equal to zero or one:
\begin{align*}
	U^{\pi}\ket{e_{\mathbf{x}}}=\ket{e_{\pi(\mathbf{x})}}
\end{align*}
and hence:
\begin{align*}
	\bra{e_{\mathbf{y}}}U^{\pi}\ket{e_{\mathbf{x}}}=1 \text{  if } \pi(\mathbf{x})=\mathbf{y} \text{  and } 0 \text{  otherwise  }
\end{align*}
Define the set $\Omega_{A}:=\{\nu_{\pi}:\pi\in A\}$ for some $A\subset S_{n}$. The point $p:=\frac{1}{n!}\sum_{\pi\in S_{n}}\nu_{\pi}$ is in the convex hull of $\Omega_{S_{n}}$:
\begin{equation}
	p\in\text{  conv  }(\Omega_{S_{n}})
\end{equation} 
where 
$$\text{  conv  }(\Omega_{S_{n}}):=\{\sum_{\pi\in S_{n}}\alpha_{\pi}\ket{\nu_{\pi}}:\forall \alpha_{\pi}\geq 0, \sum_{\pi\in S_{n}}\alpha_{\pi}=1\}$$
At this point, we can apply the Carath\'{e}odory's theorem stated above, to complete our proof. We observe that $\Omega_{S_{n}}\subset{\mathbb{R}^{d}}^{4n}$, and hence by Carath\'{e}odory's theorem, there exists a subset $X\subset S_{n}$ such that $p\in\text{conv}(\Omega_{X})$ and $|X|\leq d^{4n}+1$. Therefore there exists a weight function $\omega$ on $X$ such that:
\begin{equation}
	\sum_{\pi\in X}\omega(\pi)\nu_{\pi}=p
\end{equation}
which fulfills (\ref{conditiondesign}).
\end{proof}
The above stated bound can be also formulated in terms of entropies. Let $H(p)$ denote the Shannon entropy of the probability distribution $p$ on an alphabet $\cX$, i.e. 
\begin{align*}
 H(p) := - \sum_{x \in \cX} p(x) \log p(x),
\end{align*}
where we use the convention that $\log$ denotes base 2 logarithms. The cardinality bound in Theorem \ref{thm:symmetric_design_upperbound} implies, that we find a weighted symmetric design $(X,\omega)$ with 
\begin{align}
\frac{1}{n}H(\omega) \leq 4 \log (d+1)  \label{entropy_upper_bound}
\end{align}
 Next, we will prove a lower bound on the Shannon entropy of symmetric designs which complements the upper bound from Eq. (\ref{entropy_upper_bound}).
\begin{remark}
 The weight function $\omega$ over $X$ as defined in Def.\ref{def:weighted_symmetric_design} is a special case of a probability distribution over $S_{n}$. This can be observed by setting $\omega(\pi)=0$ for all $\pi\notin X$. When dealing with entropies, we consider $\omega$ to be a probability distribution over $S_{n}$, and hence derive a lower bound on entropy of any convex combination of permutation unitaries that produces the desired average over the group. 
\end{remark}
In what follows, we set $\cU_{\pi}(\cdot) := U_{\pi} (\cdot) U_{\pi}^\dagger$ $(\pi \in S_n)$.
\begin{theorem}\label{entropy-lowerbound}
	Let $(X,\omega)$ be a symmetric weighted design. Then:
	\begin{equation}
	\frac{1}{n}H(\omega)\geq\log(d)-2d\frac{\log(n+1)}{n}
	\end{equation}
	where $H(\omega)$ is the Shannon entropy of the weight function.
\end{theorem}
\begin{proposition}[Almost-convexity of the von Neumann entropy] \label{prop:entropy_almost_convexity}
	Let $p$ be a probability distribution on $\cX$, $|\cX| < \infty$, $\rho_x$ be a density matrix on $\cH$ where $\dim(\cH)=d$, for each $x \in \cX$, and set $\overline{\rho}_{p} := \sum_{x \in \cX} p(x) \rho_{x}$. It holds
	\begin{align}
	S(\overline{\rho}) \leq \sum_{x \in \cX} p(x) S(\rho_{x}) + H(p).
	\end{align}
\end{proposition}
\begin{proof}
	See e.g. \cite{nielsen00}, Theorem 11.10.
\end{proof}
\begin{proof}[Proof of Theorem\ref{entropy-lowerbound}]
	Fix $n \in \bbmN$, and set $\cX := \{1,\dots,d\}$ and let $\mu$ be a type of sequences in $\cX^{n}$ with
	\begin{align}\label{lemma:weight_lower_bound_3}
	H(\mu) \geq \log d - d \frac{\log(n+1)}{n}. 
	\end{align}
	Notice that existence of such a type is guaranteed by Lemma \ref{lemma:typetbound}(see Appendix\ref{appendix:types}, where more definitions and statements on frequency typical sets can also be found). Define the projection $p_\mu$
	by 
	\begin{align*}
	p_{\mu} := \sum_{\mathbf{x} \in T_{\mu}^n} \ket{e_{\mathbf{x}}}\bra{e_{\mathbf{x}}}.
	\end{align*}
	First, we notice, that for each $\mu$-typical word $\mathbf{x}$, 
	\begin{align}
	\frac{1}{n!} \sum_{\pi \in S_n} \ \cU_\pi(\ket{e_{\mathbf{x}}}\bra{e_{\mathbf{x}}}) = \frac{1}{|T_{\mu}^n|} p_\mu \label{lemma:weight_lower_bound_2}
	\end{align}
	holds (on the r.h.s. of the above equality, we find the maximally mixed state on the subspace of $\cH^{\otimes n}$ belonging to the type class $T_\mu^n$). 
	We fix a $\mu$-typical word $\mathbf{x}$ and set $E: \ket{e_{\mathbf{x}}} \bra{e_{\mathbf{x}}}$. We can bound the Shannon entropy of 
	$\omega$ by 
	\begin{align*}
	H(\omega) & \geq S\left(\sum_{\pi \in S_n} \omega(\pi) \cU_\pi(E)\right) - \sum_{\pi \in S_n} \ \omega(\pi) S(\cU_\pi(E)) \\
	& = S\left(\frac{1}{n!}\sum_{\pi \in S_n}  \cU_\pi(E)\right) \\
	& = S\left(\frac{1}{|T_\mu^n|}p_\mu\right) \\
	& = \log|T_\mu^n|.
	\end{align*}
	The inequality above is by Proposition \ref{prop:entropy_almost_convexity}. The first equality is by the fact, that $\cU_{\pi}(E)$ is a pure state
	for each $\pi \in S_n$ combined with the hypothesis of the lemma that $(X,\omega)$ is a weighted design defined by (\ref{definition}). The second equality is by (\ref{lemma:weight_lower_bound_2}). 
	We conclude
	\begin{align*}
	H(\omega) \geq n \cdot H(\mu) - \log(n+1)^d \geq n \cdot H(\mu) - 2 \cdot \log(n+1)^d.
	\end{align*}
	The left inequality above is by the standard type bound
	\begin{align*}
	|T_\mu^n| \geq \frac{1}{(n+1)^d} \cdot 2^{nH(\mu)},
	\end{align*}
	while the second is by choice of $\mu$, i.e. by the bound from (\ref{lemma:weight_lower_bound_3}). We are done.
\end{proof}
The above reasoning can be extended to derive a bound for averages of permutations which approximately simulate the action of the uniform average over $S_n$. To formulate such an assertion, we use the diamond norm $\|\cdot\|_{\diamond}$ on the set of quantum
channels on a Hilbert space $\cK$. We define
\begin{align}
 \|\cN\|_{\diamond} = \sup_{n \in \bbmN} \underset{\substack{a \in \cL( \bbmC^n \otimes \cK) \\ \|a\|_1 = 1}}{\max} \|  \id_{\bbmC^n}\otimes\cN(a) \|_1
\end{align}
for each c.p.t.p. map on $\cK$. We define c.p.t.p. maps
\begin{align}
 \overline{\cU}(b) := \frac{1}{n!} \sum_{\pi \in S_n} \cU_\pi(b) &&(b \in \cL(\cH^{\otimes n})),
\end{align}
and
\begin{align}
 \overline{\cU}_q(b) := \sum_{\pi \in S_n} q(\pi)\cU_\pi(b) &&(b \in \cL(\cH^{\otimes n}))
\end{align}
for each probability distribution $q$ on $S_n$. We prove
\begin{theorem}\label{entropy-lowerbound_approx}
	It holds
	\begin{equation}
	\frac{1}{n}H(q)\geq\log(d)-2d\frac{\log(n+1)}{n} - \tfrac{1}{n}f(\|\overline{\cU} - \overline{\cU}_q \|_{\diamond})
	\end{equation}
	for each probability distribution $q$ on $S_n$, where $\frac{1}{n}f(x) \rightarrow 0$, $(x \rightarrow 0)$. More specifically, $f(x):=2x\log(d-1)+2H_{2}(x)$ where $H_{2}(x)$ is the binary entropy and $d$ is the dimension of the underlying Hilbert space.
\end{theorem}
\begin{proof}
 The proof is by minor extension of the argument given to prove Theorem \ref{entropy-lowerbound}. Note, that with $E := \ket{e_{\mathbf{x}}} \bra{e_{\mathbf{x}}}$ as in the proof of Theorem \ref{entropy-lowerbound}
 \begin{align}
  \epsilon := \|\overline{\cU} - \overline{\cU}_q\|_\diamond \geq \|(\overline{\cU} - \overline{\cU}_q)(E)\|_1
 \end{align}
 holds. By a sharp version of Fannes' inequality due to Audenaert (\cite{fannes73}), we have
 \begin{align}
  S(\overline{\cU}_q) \geq S(\overline{\cU}) - f(\epsilon) 
 \end{align}
 with a function fulfilling $\frac{1}{n}f(\epsilon) \rightarrow 0$ ($\epsilon \rightarrow 0$). We repeat the line of reasoning from the proof of Theorem \ref{entropy-lowerbound} including the above tradeoff to the inequalities and get
 \begin{align}
  H(q) 
  &\geq S(\overline{\cU}_q(E)) \\
  &\geq S(\overline{\cU}(E)) - f(\epsilon) \\
  &\geq n \log d - 2 d (n+1) - f(\epsilon).
 \end{align}
\end{proof}
The bounds obtained so far directly imply corresponding bounds for completely positive and trace preserving (c.p.t.p.) matrices.
\begin{theorem} \label{lowerbound_channels}
  Let $\dim \cK := d_\cK$, $\dim \cH := d_\cH$, and $\cU_\pi(\cdot) := U^\pi(\cdot)(U^{\pi})^\ast$, $\cV_\pi(\cdot) := V^\pi(\cdot)(V^{\pi})^\ast$ 
  be the c.p.t.p. maps permuting the tensor factors on $\cL(\cH)^{\otimes n}$ resp. $\cL(\cK)^{\otimes n}$ according to $\pi$ for each $\pi \in S_n$. If
  \begin{align}
  \frac{1}{n!} \sum_{\pi \in S_n} \cU_{\pi} \circ \cN \circ \cV_{\pi^{-1}} = \sum_{\pi \in S_n} \omega(\pi) \cU_{\pi} \circ \cN \circ \cV_{\pi^{-1}}
 \end{align} 
 for each c.p.t.p. map $\cN: \cL(\cH^{\otimes n}) \rightarrow \cL(\cK^{\otimes n})$, then
 \begin{align}
  \frac{1}{n}H(\omega)\geq\log(d_\cK d_{\cH}) -2d_\cK d_\cH \frac{\log(n+1)}{n}
 \end{align}
\end{theorem}
 \begin{proof}
  The proof of the above assertion almost immediately follows from Theorem \ref{entropy-lowerbound} combined with the Jamio\l kowski isomorphism (see e.g. \cite{wilde13})
  \begin{align}
   \cN \mapsto \sigma_\cN := \cN \otimes \id_{\cH^{\otimes n}}(\ket{\Phi} \bra{\Phi}),
  \end{align}
 where $\ket{\Phi}$ is defined by
 \begin{align}
  \ket{\Phi} := \frac{1}{d^n} \sum_{\mathbf{x} \in \cX^{n}} \ket{e_{\mathbf{x}}} \otimes  \ket{e_{\mathbf{x}}}
 \end{align}
 Indeed, for each c.p.t.p. map $\cN: \cL(\cH)^{\otimes n} \rightarrow \cL(\cK)^{\otimes n}$, it holds 
 \begin{align}
  \sigma_{\cU_{\pi} \circ \cN  \circ \cV_{\pi^{-1}}} 
  &= \cU_{\pi} \circ \cN  \circ \cV_{\pi^{-1}} \otimes \id_{\cH^{\otimes n}}(\ket{\Phi} \bra{\Phi}) \\
  &= \cU_\pi \otimes \cV_\pi(\sigma_{\cN}).
 \end{align}
 \end{proof}

A lower bound on the cardinality of designs (and 2-designs by a straightforward extension) can be readily established from Theorem \ref{entropy-lowerbound}. We finish this section, however, by remarking a relation between vectors belonging to the symmetric subspace and permutation invariant states, that in turn enables us to derive a lower bound on the cardinality of designs.\\ 

It can be observed that permutation invariant matrices are not in general supported on $\text{sym}^{(n)}(\mathcal{H})$, the subspace defined by:
\begin{equation*}
\text{sym}^{(n)}(\mathcal{H}):=\text{span}(\ket{\nu}:U^{\pi}\ket{\nu}=\ket{\nu}\forall\pi\in S_{n})
\end{equation*}
An example to the point is $M=\ket{e_{01}}\bra{e_{01}}+\ket{e_{10}}\bra{e_{10}}$ where $\ket{e_{ij}}=\ket{e_{i}}\otimes\ket{e_{j}}$. The following lemma from \cite{phdthesis} can be used to establish a relation between permutation invariant states and vectors on $\text{sym}^{(n)}(\mathcal{H})$:
\begin{lemma}[\cite{phdthesis}, Lemma 4.2.2]\label{corollary}
Let the state $\rho_{n}\in S(\mathcal{H}^{\otimes n})$ be permutation invariant and have the following spectral decomposition:
\begin{equation*}
    \rho_{n}:=\sum_{i}\lambda_{i}\ket{\nu_{i}}\bra{\nu_{i}}
\end{equation*}
where we have included the zero eigenvalues. Then $\ket{\psi}:=\sum_{i}\sqrt{\lambda_{i}}\ket{\nu_{i}}\otimes\ket{\nu_{i}}\in\text{sym}^{(n)}(\mathcal{H})$.
\end{lemma}
Using this lemma, we prove a lower bound on the cardinality of symmetric weighted designs:
\begin{theorem}
	Let $(X,\omega)$ be a weighted design for $S_{n}$. Then we have:
	\begin{equation}
	|X|\geq d^{n}-\binom{d+n-1}{d-1}
	\end{equation}
\end{theorem}
\begin{proof}
	Consider $\ket{\nu}\in\text{sym}^{(n)}(\mathcal{H})^{\perp}$, where the superscript indicates the orthogonal compliment. It can be observed that $U^{\pi}\ket{\nu}\in\text{sym}^{(n)}(\mathcal{H})^{\perp}\forall \pi\in S_{n}$. To see this, we notice that $\forall\ket{\psi}\in\text{sym}^{(n)}(\mathcal{H})$ we have $\bra{\psi}U^{\pi}\ket{\nu}=\braket{\psi|\nu}=0$. The second equality is due to the fact that $\ket{\psi}$ is permutation invariant and absorbs $U^{\pi}$. Consider the set $V:=\{U^{\pi}\ket{\nu}\}_{\pi\in X}$ for some $X\subset S_{n}$. If $|X|<\dim(\text{sym}^{(n)}(\mathcal{H})^{\perp})$,we can orthonormalize this set via Gram-Schmidt process and obtain $V^{\prime}:=\{\ket{\nu^{\pi}}\}_{\pi\in X}$. $V^{\prime}$ would then be an ONB for a subspace of $\text{sym}^{(n)}(\mathcal{H})^{\perp}$. Finally, define $\tilde{V}:=\{ \ket{\nu^{\pi}}\otimes\ket{\nu^{\pi}}\}_{\pi\in X}$. It can be observed that $\ket{\nu^{\pi}}\otimes\ket{\nu^{\pi}}\in\text{sym}^{(n)}(\mathcal{H}\otimes\mathcal{H})^{\perp}$. 
	There are two possibilities for any linear combination with non-zero multiples of elements in $\tilde{V}$: for any set $\{\lambda^{\pi}\neq0,\pi\in X\}$ either:
	\begin{equation*}
	1. \sum_{\pi\in X}\lambda^{\pi}\ket{\nu^{\pi}}\otimes\ket{\nu^{\pi}}=0
	\end{equation*}
	or
	\begin{equation*}
	2.\sum_{\pi\in X}\lambda^{\pi}\ket{\nu^{\pi}}\otimes\ket{\nu^{\pi}}\neq 0 \text{  and  }\in\text{sym}^{(n)}(\mathcal{H}\otimes\mathcal{H})^{\perp}
	\end{equation*}
	But the first case cannot be, as $\{\ket{\nu^{\pi}}\otimes\ket{\nu^{\pi}}\}_{\pi\in X}$ is linearly independent for $|X|<\dim(\text{sym}^{(n)}(\mathcal{H})^{\perp})$. The second case, by Lemma \ref{corollary} implies that the state $\sigma:=\sum_{\pi\in X}(\lambda^{\pi})^{2}\ket{\nu^{\pi}}\bra{\nu^{\pi}}$ cannot be permutation invariant. Since $\lambda^{\pi}$ is any non-zero number, this is true for all linear combinations of states $\ket{\nu^{\pi}}\bra{\nu^{\pi}}$ as long as $|X|<\dim(\text{sym}^{(n)}(\mathcal{H})^{\perp})$. But what does this imply for linear combinations of states $U^{\pi}\ket{\nu}\bra{\nu}{U^{\pi}}^{\dagger}$ for $\pi\in X$. For any such state $$\mu:=\sum_{\pi\in X}\omega^{\pi}U^{\pi}\ket{\nu}\bra{\nu}{U^{\pi}}^{\dagger}$$
	we have:
	\begin{equation*}
	    \mu=\sum_{\pi\tilde{\pi}\in X}\gamma^{\pi\tilde{\pi}}\ket{\nu^{\pi}}\bra{\nu^{\tilde{\pi}}}
	\end{equation*}
	In the ONB given by $V^{\prime}$, the right hand side can be decomposed into a diagonal matrix $Q$ and an off-diagonal matrix $R$. The diagonal matrix is a linear combination of states $\ket{\nu^{\pi}}\bra{\nu^{\pi}}$ and hence cannot be permutation invariant by arguments given above. But for $\mu=Q+R$ to be permutation invariant, both $Q$ and $R$ have to be permutation invariant, as application of any unitary on $\mu$ will produce a diagonal matrix and an off-diagonal one, cancelling out $Q$ and $R$ respectively when considering $\mu-U^{\pi}\mu {U^{\pi}}^{\dagger}$.
\end{proof}

\section{Communication-theoretic implications of the results} \label{sect:applications}
\begin{figure}
\begin{center}
\includegraphics[width=.9\linewidth]{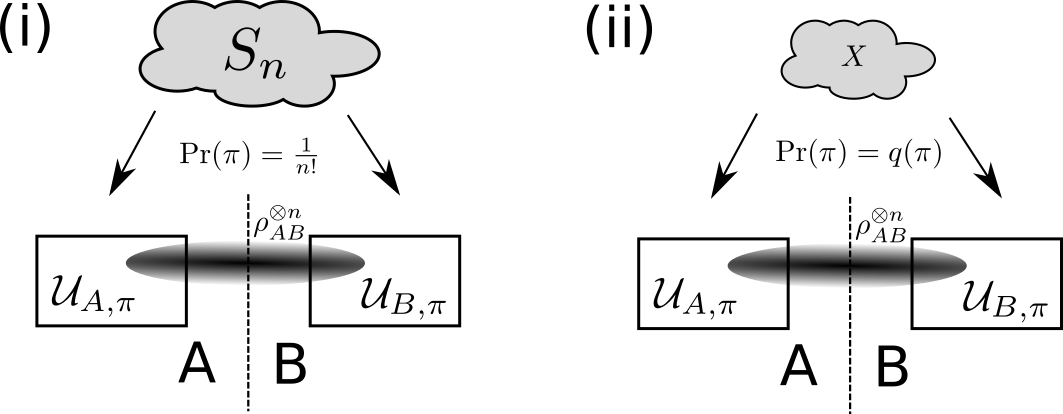} \vspace{2ex}  
\caption{(i) Implementation of $\overline{\cU}$ by equidistributed and correlated random choice of permutations. (ii) Simulation of $\overline{\cU}$ by correlated choice of a random permutation from a smaller set $X$ according to probability distribution $q$} \label{bild_1}
\end{center}
\end{figure}

In this section, we discuss some consequences of the technical results from the the last section. From the upper and lower bounds derived there, some remarkable conceptual implications in communication theory can be drawn. \\
Assuming $\cH$ as the underlying Hilbert space of the system under consideration, the 
quantum channel
\begin{align*}
 \overline{\cU}(\cdot) := \frac{1}{n!} \sum_{\pi \in S_n} \cU_\pi(\cdot), \hspace{.3cm}
 \cU_{\pi}(\cdot) := U^{\pi} (\cdot) {U^{\pi}}^\dagger &&(\pi \in S_n)
\end{align*}
is usually regarded as the standard protocol applied to universally map each state 
on $\cH^{\otimes n}$ to a permutation invariant one. \footnote{In this section, we restrict ourselves to discussion of the consequences of the derived bounds for quantum states. Similar observation regarding quantum channels easily follow from our bounds
regarding quantum channels.} \\
To zest the discussion, we consider $\cH := \cH_A \otimes \cH_B$ the space of a bipartite system shared by distant communication parties $A$ and $B$. The corresponding map $\overline{\cU}$ on $\cH$ has the form
\begin{align}
 \overline{\cU}(\cdot) = \frac{1}{n!} \sum_{\pi \in S_n} \cU_{A,\pi} \otimes \cU_{B,\pi}(\cdot),
\end{align}
where $\cU_{A,\pi}$ and $\cU_{B,\pi}$ are the channels exchanging the subsystems of $\cH_A^{\otimes n}$ respectively $\cH_B^{\otimes n}$ according to permutation $\pi$. 
To implement $\overline{\cU}$ as a communication protocol, $A$ and $B$ have to agree on a permutation
which is chosen randomly from the symmetric group $S_n$ on $n$ letters (see Figure \ref{bild_1}). 

\begin{figure}
\begin{center}
\includegraphics[width=.8\linewidth]{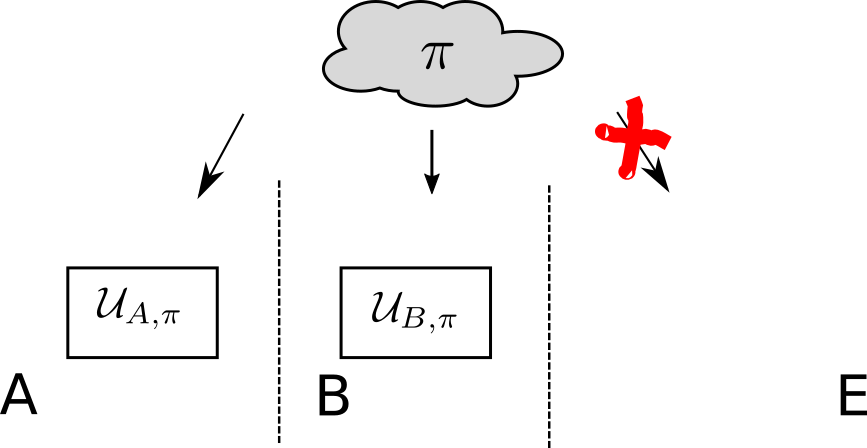} \vspace{2ex}  
\end{center}
\caption{Choice of random permutation $\pi$ for implementation of $\overline{\cU}$ has to be coordinated between legitimate parties $A$ and $B$ but protected from knowledge by adversarial party $E$.} \label{bild_2}
\end{figure}

Applying $\cU$ as a communication protocol (or as a part of a greater protocol) consequently amounts in consuming shared equidistributed randomness (\emph{common 
randomness} as it is called usually in the information theory literature) at rate
\begin{align}
 R_n = \frac{1}{n} \log n!
\end{align}
bits per block length. The observation, that the rates $R_n$ grow unbounded in the asymptotic limit $n \rightarrow 
\infty$ disqualifies $\cU$ as a protocol in situations, where shared randomness is not a free resource, but instead does count to the resource trade-off. \\
In this context, Theorem \ref{thm:symmetric_design_upperbound} proven in the preceding 
section provides an uplifting message. A weighted symmetric 
design (on $\cH^{\otimes n}$) as introduced in Definition \ref{def:weighted_symmetric_design} exactly simulates the action of $\overline{\cU}$.
Theorem \ref{thm:symmetric_design_upperbound} therefore shows, that we always can equivalently replace $\overline{\cU}$ by a protocol which demands (not necessarily 
equidistributed randomness) at a rate 
\begin{align}
 R'_n \leq 4 \cdot \log \dim(\cH_A \otimes \cH_B). \label{sect_3_1}
\end{align}
We have shown, that the brute-force evenly distributed random selection out of all permutations can be replaced by random selection from a much smaller set of permutations (which amounts to rate-bounded coordinated randomness demands.) \\
Opposite to the consequences discussed so far, our results also enforce some conclusions of the more disillusioning type. Having established protocols for enforcing permutation-invariance which are reasonable regarding their randomness consumption, they may be too expensive in randomness consumption sometimes. \\
As a consequence of the well-known Holevo bound, we obtain the inequality 
\begin{align}
  I(X_{A^{n}};Y_{B^{n}}) \ \leq \ n \cdot \log \dim \cH_A \otimes \cH_B \label{sect_3_2}
\end{align}
which provides a principal bound for the mutual information of a bipartite random variable $(X_{A^{n}}, Y_{B^{n}})$
produced by local measurements on the $A$ and $B$ subsystems of any bipartite 
quantum system with underlying Hilbert space $(\cH_A \otimes \cH_B)^{\otimes n}$. 
When regarding resource trade-offs, comparing the bounds in (\ref{sect_3_2}) and the one given by 
\begin{align}
 H(p) \ \geq \ n \log \dim \cH_A \otimes \cH_B
\end{align} 
for Shannon entropy of any probability distribution producing a symmetric design given by
\begin{align}
 \overline{\cU}_{q} \ := \sum_{\pi \in S_n}  \ q(\pi) \cdot \cU_{A,\pi} \otimes \cU_{B,\pi} 
\end{align} 
we notice that permutation-symmetrization, costs at least as much shared randomness as could be produced at all (in a perfect situation) by local measurements on a system.\\
While the preceding observation may have no consequences in communication situations where shared randomness is a cheap resource, there are other situations, where the communication demands are critical to an extent, that the introduced protocol class is disqualified. \\
A special instance of such a situation is faced, when in addition to $A$ and $B$ (which we call henceforth \emph{legitimate users}) a third, malicious party $E$ takes part in the communication. Let the 
underlying space of the system be $\cH_A \otimes \cH_B \otimes \cH_E$. In this case, it is usually not enough to perform the random choice in a way that it is coordinated between the legitimate parties $A$ and $B$. Moreover, it has to be secure in the sense, that the malicious party $E$ has no knowledge of the permutation $\pi$ chosen (see Figure \ref{bild_2}.)  \\
An example where the correlation shared not just by the legitimate but also the adversarial communication parties is useless, is given in \cite{boche17}. Therein it is proven, that
the secrecy capacity of an arbitrarily varying wiretap classical quantum channel (AVWQC) under assistance of common randomness which is secure against the jamming adversarial party sometimes strictly exceeds the corresponding capacity of the AVWC under assistance of public (non-private) common randomness. Additionally, in the case where the randomness is also accessible to the jamming adversary, the corresponding capacity equals the capacity without any common randomness assistance. Common randomness is useless for secret message transmission if it is also known to the (active) adversary.\\
\section{Conclusion} \label{sect:discussion}
In this paper, we derived upper and lower bounds of randomness consumption of universal symmetrization of states by applying averaged permutations. Our bounds lead to positive and negative conclusions when applied to information-theoretic
modelling. First, the encouraging implication of our upper bound on the support of weighted designs shows, that there are always protocols which universally symmetrize quantum states on a given $n$-partite quantum system, 
consuming reasonable common randomness resource. Specifically, the number of coordinated random choices of permutations used for symmetrizing arbitrary quantum states on that system can be always restricted to being exponentially growing (with number of systems). The lower bounds on the common randomness needed for permutation-based symmetrization of arbitrary quantum states proven in this paper, enforce a rather disillusioning conclusion. To universally symmetrize 
quantum states on the $n$-fold tensor extension of a given system with Hilbert space dimension $d$, one asymptotically needs at least a common randomness rate of $\log d$. Since this number marks the trivial upper bound for common randomness 
rates generated from repetitions of an ideal system of that dimensionality, the common randomness consumption seems exorbitant in some situations. \\
Since universal symmetrization of communication attacks is a vital ingredient of a broad class of security proofs for quantum key distribution protocols, our findings strongly motivate further research for finding more efficient protocols 
for symmetrizing quantum states and channels.
\begin{section}*{Acknowledgements}
This work was supported by the BMBF via grant 16KIS0118K.
 \end{section}

 \begin{section}*{Appendix: Frequency typical sets} \label{appendix:types} In this appendix we give the basic definitions regarding types and frequency typical sets. For a broad as well as concise introduction to the concept of types the reader is referred to \cite{csiszar11}, where the bounds stated in this appendix can be found without exception. \\ Let $\cX$ be a finite set, $p$ a probability distribution on $\cX$. We define the set of $p$-typical words in $\cX^{n}$ by \begin{align*}
	T_p^n := \{\mathbf{x}: \ \forall a \in \cX: \tfrac{1}{n}N(a|\mathbf{x}) = p(a) \}.
	\end{align*}
	If this set in nonempty, we call $p$ a \emph{type of sequences in $\cX^{n}$} (or \emph{n-type} for short). The concept of types is a powerful tool in classical as well as quantum Shannon theory. In this paper, use some cardinality bounds on the entities introduced which are stated in the next two lemmas. If we denote, for $n \in \bbmN$ the set of $n$-types by $\fT(n,\cX)$, the following statement is true.
	\begin{lemma}[cf. \cite{csiszar11}, Lemma 2.2] \label{appendix:type-bound_1}
		For each $n \in \bbmN$, it holds
		\begin{align*}
		|\fT(n,\cX)| \leq (n+1)^{|\cX|}.
		\end{align*}
	\end{lemma}
	
	\begin{lemma}[cf. \cite{csiszar11}, Lemma 2.3]
		For each $n \in \bbmN$, and each $n$-type $\lambda \in \fT(n,\cX)$, it holds
		\begin{align*}
		(n+1)^{-|\cX|} \cdot 2^{nH(\lambda)} \ \leq \ |T_{\lambda}^n| \ \leq 2^{nH(\lambda)}.
		\end{align*}
		
	\end{lemma}
\begin{lemma}\label{lemma:typetbound}
	For each $n\in \bbmN$, there exists a type $\mu_\ast$ of sequences in $\cX^{n}$, such that
	\begin{align}
	H(\mu_\ast) \geq \log|\cX| - |\cX| \frac{\log(n+1)}{n} \label{lemma:typetbound_1}
	\end{align}
	holds.
\end{lemma}

\begin{proof}
	Let $\mu_\ast$ be a type of sequences in $\cX^{n}$, which maximizes the Shannon entropy, i.e.
	\begin{align*}
	H(\mu_\ast) \geq H(\lambda)
	\end{align*}
	holds for each type $\lambda$. Then, by standard bounds for the frequency typical sets \cite{csiszar11}
	\begin{align*}
	|T_{\lambda}^n| \leq 2^{nH(\lambda)} \leq 2^{n H(\mu_\ast)}. 
	\end{align*}
	holds for each type $\lambda$. Since there are not more than $(n+1)^{|\cX|}$ different types of sequences in $\cX^{n}$,
	the bound 
	\begin{align*}
	|\cX^{n}| \leq (n+1)^{d} \cdot 2^{nH(\mu_\ast)}
	\end{align*}
	is valid, which with some rearrangements proves the lemma. 
\end{proof}
	\end{section}


\begin{thebibliography}{10}
\bibitem{ahlswede78}
R. Ahlswede. 
\newblock Elimination of correlation in random codes for arbitrarily varying channels.
\newblock {\em Z. f. Wahrsch. Th.} 44, 159--175 (1978).



\bibitem{fannes73}
K.M.R. Audenaert.
\newblock  A Sharp Fannes-type Inequality for the von Neumann Entropy.
\newblock {\em J.Phys.A.} 40, 8127-8136 (2007).
\bibitem{barvinok03}
A. Barvinok. 
\newblock {\em A course in convexity},
\newblock Graduate studies in mathematics 54. American Mathematical Society (2003)

\bibitem{bb84}
C.H. Bennett, G. Brassard.
\newblock Quantum cryptography: Public key distribution and coin tossing. 
\newblock {\em Proc. IEEE Int. Conference on Computers Systems and Signal Processing} 175--179 (1984). 

\bibitem{bennett93}
C.H. Bennett and G. Brassard and C. Cr\'{e}peau and R. Jozsa and A Peres and W.K. Wootters.
\newblock Teleporting an unknown quantum state via dual classical and Einstein-Podolsky-Rosen channels.
\newblock {\em Phys. Rev. Lett.} 70, 1895--1899 (1993).

\bibitem{boche17}
H. Boche, M. Cai, C. Deppe, J. N\"otzel.
\newblock Classical-Quantum Arbitrarily Varying Wiretap
Channel: Secret Message Transmission under
Jamming Attacks
\newblock{\em To appear in proceedings of ISIT} (2017).

\bibitem{csiszar11}
I. Csisz\'{a}r, J. K\"{o}rner.
\newblock {\em Information Theory - Coding Theorems for Discrete Memoryless Systems},
\newblock 2nd Ed. Cambridge University Press (2011).


\bibitem{nielsen00}
M. A. Nielsen, I. L. Chuang.
\newblock {\em Quantum Information and Computation}.
\newblock Cambridge University Press (2000). 

\bibitem{webster94}
R.~Webster.
\newblock{\em Convexity}.
\newblock Oxford University Press, 1994.

\bibitem{wilde13}
M.~Wilde.
\newblock {\em Quantum Information Theory}.
\newblock Cambridge University Press (2013). 

\bibitem{wakakuwa16}
E. Wakakuwa. 
\newblock{Symmetrizing Cost of Quantum States.}
\newblock{\em Phys. Rev. A 95}, 032328 (2017).

\bibitem{als} M.Christandl, R.K\"{o}nig, R.Renner, "Post-selection technique for quantum channels with applications to quantum cryptography ",  Physical Review Letters 102(2):020504  (2009).
\bibitem{cara} H.G. Eggleston, "Convexity", Cambridge University Press. doi:10.1017/cbo9780511566172 (1958). 
\bibitem{evenly} D. Gross, K. Audenaert, and J. Eisert, "Evenly distributed unitaries: On the structure of unitary designs", Journal of Mathematical Physics (2007).
\bibitem{harrow}A. W. Harrow, " The church of the symmetric subspace", arXiv preprint arXiv:1308.6595, (2013).
\bibitem{winter} S. Karumanchi, S. Mancini, A. Winter and D. Yang, "Quantum Channel Capacities
with Passive Environment Assistance ", arXiv[quant-ph]:1407.8160 (2014).
\bibitem{konig} H.K\"{o}nig, "Cubature Formulas on Spheres ", Adv. Multivar. Approx., Math. Res. 107 (1999), Wiley, 201 - 211.
\bibitem{phdthesis} R.Renner "Security of Quantum Key Distribution, PhD Thesis",  arXiv:quant-ph/0512258v2, (2005).

\end{thebibliography}
\end{document}